\documentclass[twocolumn,letterpaper,aps,prc,superscriptaddress,nofootinbib,showpacs,floatfix]{revtex4-1}
\usepackage{graphicx}
\usepackage{comment}
\usepackage{setspace}
\usepackage{amsmath}
\usepackage{amssymb}
\usepackage[textwidth=16cm,textheight=22cm]{geometry}
\usepackage{color}
\usepackage{indentfirst}
\usepackage{xspace}
\usepackage{hyperref}
\usepackage{verbatim}
\usepackage{epstopdf}

\usepackage{lineno}
{\large }

\begin{document}
\title{Understanding long-range near-side ridge correlations in p$-$p collisions using rope hadronization at LHC energies.}

\author{Pritam Chakraborty}
\email{prchakra@iitb.ac.in}
\affiliation{Indian Institute of Technology Bombay, Mumbai, 
  India-400076}

\author{ Sadhana~Dash }
\email{sadhana@phy.iitb.ac.in}
\affiliation{Indian Institute of Technology Bombay, Mumbai, 
  India-400076}

\begin{abstract}
The observation of long range ridge-like structure in the near-side region of the two particle $\Delta\eta-\Delta\phi$ correlations as 
measured by LHC experiments in high multiplicity p$-$p collisions indicated towards the presence of collective effects  which are similar to that 
observed in p$-$A(nucleon-nucleus) and A$-$A (nucleus-nucleus) collisions. The
two particle correlation between the charged particles in $\Delta\eta-\Delta\phi$ 
for p$-$p collisions at $\sqrt{s}$ = 7 TeV  and 13 TeV is studied using Pythia 8 event
generator within the framework of final-state partonic color reconnection
effects as well as the microscopic rope hadronization model. The rope
hadronization relies on the formation of ropes due to overlapping of
strings in high multiplicity events followed by string shoving.  A  near side ridge-like structure which is qualitatively similar to the observed ridge 
in data was observed for high-multiplicity events when the  mechanism of rope hadronization (with shoving) was enabled. 
 \end{abstract}

\maketitle

\section{Introduction}

The recent observation of long-range ridge like structure in the near side region of  $\Delta\eta-\Delta\phi$ correlations of charged particle pairs measured by CMS and ATLAS experiment at LHC  \cite{cms1,cms2,atlas} have generated a lot of interest in small systems. An enhanced correlation was found for charged particle pairs which are highly collimated in azimuthal angle over large $\eta$ intervals. These ridge-like structures and long range correlations were not previously observed in minimum bias events of hadronic collisions. Such observations are reminiscent of ridge-like features present 
in heavy ion collisions and are believed to be manifestations of hydrodynamic collectivity in partonic matter\cite{rhic2part,lhc2part}. However, it is also expected 
that no-QGP like effects should be seen in p$-$p collisions as the size of system formed in hadronic collisions is extremely small and short lived. As the particle production in high multiplicity events in hadronic collisions are highly influenced by non-perturbative QCD processes, the study of  such events can provide information  about the potential microscopic processes leading to such novel observations. 

The CMS experiment at LHC observed a non-trivial ridge-like structure for the first time in the near side, long range region of  the 
two-dimensional $\Delta\eta-\Delta\phi$ distributions of charged particle pairs in high multiplicity p$-$p collisions at $\sqrt{s}$ = 7 TeV  and 13 TeV \cite{cms1,cms2}. The correlation structure was found to be more pronounced  in the intermediate $p_{T}$ range and for high multiplicity events. It is well known that the two particle correlation function in minimum bias events is conspicuous with its enhanced structure due to contributions from jet fragmentation, resonance decays, string fragmentation etc at ($\Delta\eta$, $\Delta\phi$)  $\sim$ (0,0). The fragmentation of back-to-back jets contribute a broad ridge-like structure around  $\Delta\phi \sim \pi$ and the principle of momentum conservation leads to a shallow dip at $\Delta\phi = 0$ for larger $\Delta\eta$ ranges. However, in nucleus-nucleus (A$-$A) collisions, a ridge-like structure was also observed in the near-side region around $\Delta\phi = 0$ for large values of  $\Delta\eta$ for correlated pairs of hadrons as measured by RHIC and LHC experiments\cite{rhic2part,lhc2part}. The strength ( or amplitude) of these correlations increased with collision centrality. This observation of 
near-side long range rapidity correlation was attributed to the presence of strong collectivity of the matter created in heavy ion collisions\cite{collect1}.  However,  many theories related to the response of the medium to the partonic interactions, initial fluctuations in the geometry of overlapping nuclei leading to triangularity (and higher order) final state azimuthal anisotropy, etc were also investigated to understand the novel phenomenon\cite{collect2,collect3,collect4}. Previously, long range correlations were also observed in hadronic collisions at ISR experiments but they were not collimated in azimuthal angle and thus did not resemble like a ridge \cite{isr}. The LHC experiments also studied the two particle correlations in high multiplicity p$-$Pb collisions and  observed the  
striking ridge-like correlations in the near side, long range region \cite{ridgepPb}. However, the magnitude of the structure was found to be five times higher than that observed 
in p$-$p collisions.  Many alternative explanations based on hydrodynamic evolution \cite{epos1} and the formation of color glass condensate \cite{cgc1,cgc2} could explain the observation of ridge in p$-$p collisions. A recent study using the DIPSY generator \cite{dipsy} which incorporated the shoving mechanism in rope hadronization could qualitatively explain the collective phenomenon of the formation of near side 
ridge in long range correlations for p$-$p collisions\cite{dipsyproc}. However, the study was shown in a biased sample of DIPSY events where the events had only long strings. 
In this work,  the two particle correlation of charged particles in $\Delta\eta-\Delta\phi$ is investigated in details using the shoving mechanism implemented in 
Pythia 8 \cite{pythia8} event generator for p$-$p collisions at  $\sqrt{s}$ = 7 TeV and 13 TeV.  
The theory of color reconnections and string hadronization via the formation of ropes could explain the strangeness enhancement and the suppression of yields of resonance particles  
in p$-$p collisions at $\sqrt{s}$ = 7 TeV and 13 TeV using Pythia 8 \cite{sdashrope,bielrich1,bielrich2,sdashreso}. The model does not assume the formation of a deconfined  and thermalized plasma state 
due to the dynamical interactions between the partons. In p$-$p collisions at high energy, one can have multitude of partonic interactions in a single event which is more pronounced in high multiplicity events.  The various sub-processes during multi-partonic interactions (MPI) are reconnected and the colour flow is reassigned in the beam remnant by different processes which define the different mode of color reconnection mechanism. The mechanism of  color reconnections refers to color connections between the final partonic strings before hadronization.  The MPI-mode  (CR-0) refers to the fusion of colour flow so that to yield  the smallest  string length. Recently, the other modes namely the QCD mode (CR-1) and the gluon-move scheme (CR-2) were introduced which minimized the string length on basis of QCD color rules and movement of gluons \cite{color1,color2}.
In high multiplicity events, these color strings overlap with each other and act coherently to form a color rope that subsequently hadronize with a higher effective string tension \cite{rope1,rope2}. The greater energy density of the string overlap region creates a dynamic pressure gradient which push the strings in the outward direction.  The $p_{T}$, transverse momentum, of the strings is generated from the excess energy in the central region. Due to this outward movement of strings, the pressure gradient as well as the excess energy gradually decreases and consequently the strings pick less $p_{T}$  until there is no overlap. Thus, these strings shove each other picking up extra transverse boost and hence an increase of mean $p_{T}$ for heavier hadrons is observed.  This observation is similar to the that obtained in case of hydrodynamic scenario. The effects of different modes of color reconnections are also studied for both the energies.

\section{ Analysis Method }
Generally, the two particle correlations are studied in terms of two-dimensional $\Delta\eta-\Delta\phi$
correlation functions \cite{cms1,cms2}. In this work, the generated events are classified into different class of multiplicities depending 
on the number of particles produced within $|\eta| < 2.4$ and $p_{T} > 0.4$ GeV/c. For a given multiplicity class, the trigger particles are selected from a 
given $p_{T}$ range and the total number of trigger particles is referred as $N_{trig}$. Each trigger particle is then associated with the remaining charged particles (called as associated particles ) from a certain $p_{T}$ range to form particle pairs. The per trigger particle yield of such pairs from same event is S($\Delta\eta$, $\Delta\phi$) and is expressed as :
 
\begin{eqnarray}
S(\Delta\phi,\Delta\phi)=\frac{1}{N_{trig}}
\frac{d^2N^{same}}{d\Delta\eta d\Delta\phi}.   \label{eq2}
\end{eqnarray}  

The symbol $\Delta\eta$ and $\Delta\phi$ denotes the differences in $\eta$ and $\phi$ of the formed pair.
The mixed event pair distribution is constructed by forming pairs of the trigger particles in a given event with
the associated particles from different events. The number of events mixed for this analysis was 10. 
The mixed-event pair distribution is given by
\begin{eqnarray}
B(\Delta\phi,\Delta\phi)=\frac{1}{N_{trig}}
\frac{d^2N^{mix}}{d\Delta\eta d\Delta\phi}.   \label{eq3}
\end{eqnarray}

The distribution of signal pairs and background pairs are constructed event wise for a given multiplicity class and are then averaged over all considered events to get the distribution of the corrected per trigger particle associated yield. 
The corrected per trigger particle associated yield is then defined as
\begin{eqnarray}
\frac{1}{N_{trig}}\frac{d^2N^{pair}}{d\Delta\eta d\Delta\phi}=
B(0,0)\frac{S(\Delta\eta,\Delta\phi)}{B(\Delta\eta,\Delta\phi)}     \label{eq1}
\end{eqnarray}

The Pythia 8 event generator \cite{pythia8} was used to generate 40 million events for p$-$p collisions at $\sqrt{s}$  = 7 TeV and 13 TeV and 
the events were analysed to obtain the two dimensional  $\Delta\eta-\Delta\phi$ correlation functions for charged 
particles in different multiplicity classes. The trigger and associated particles have been chosen from the
$p_T$  ranges,  0.5 - 3 GeV/c and 1 - 3 GeV/c  for  three different multiplicity class ranging from $0<N_{trk}<20$ , $80<N_{trk}<100$  and $N_{trk}>100$. 
The $N_{trk}$ refers to the number of charged particles within the acceptance of 
$|\eta| < 0.5$. The Monash 2013 tune of Pythia 8.2 \cite{monash} has been used to
generate events for p$-$p collisions. The multi-partonic interactions were enabled for the study. The three different 
modes of color reconnection mechanism  (namely MPI-based(CR0), QCD -based(CR1) and gluon-move(CR2) ) were studied with (and without) the rope hadronization scenario to observe the effect of color reconnection and rope hadronization on two particle correlations. 

\section{ RESULTS }
\begin{figure*}

\begin{center}
\includegraphics[scale=0.78]{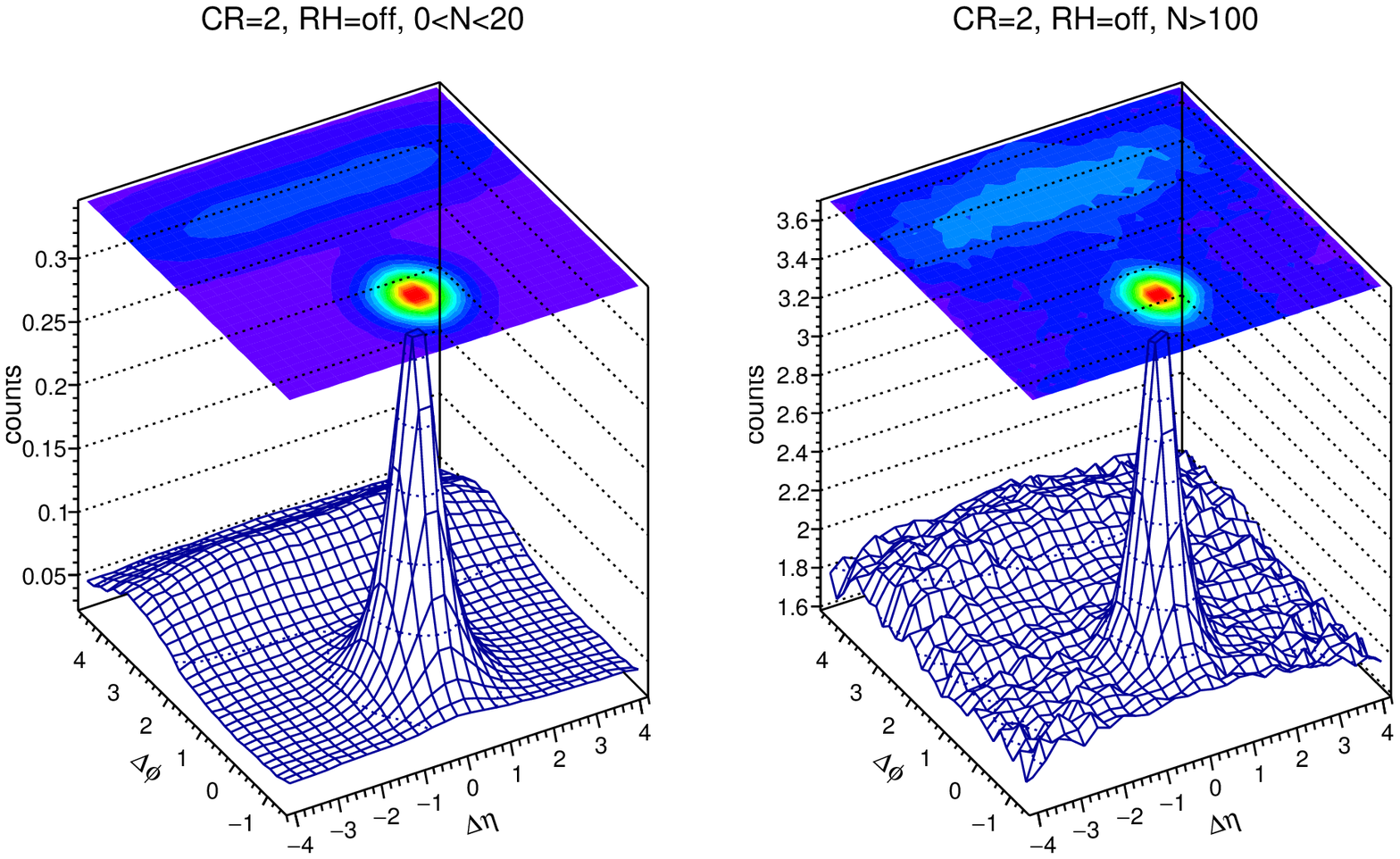}
\includegraphics[scale=0.78]{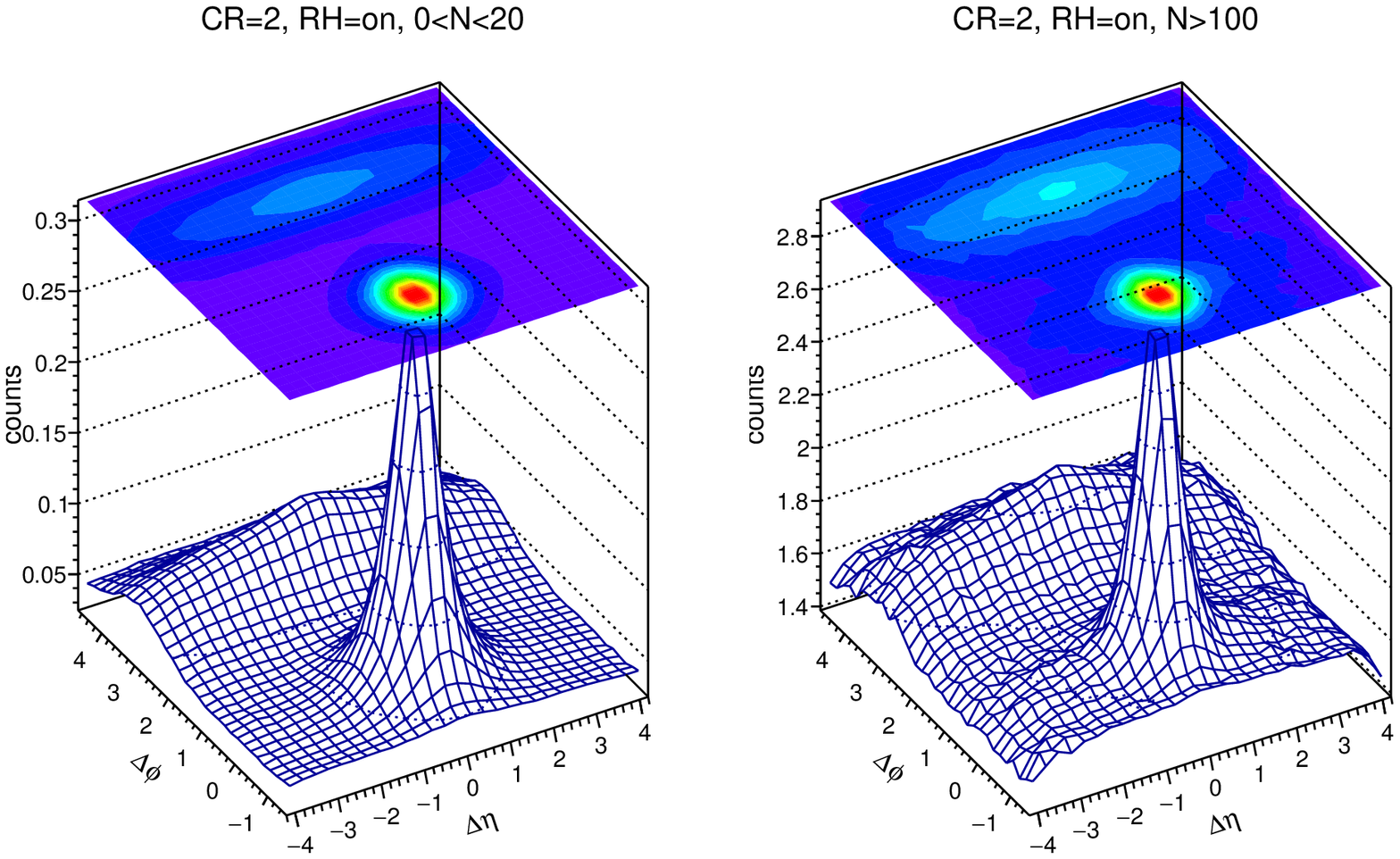}
\caption{(Color online) The two-dimensional $\Delta\eta-\Delta\phi$ correlation function of charged particle pairs in p$-$p collisions at  $\sqrt{s}$ = 7 TeV for two different multiplicity classes with $1.0 < p_{T}^{Trigg}<
3.0$  GeV/c and  $1<p_{T}(Asso)< 3.0$. The top panels refer to the case when rope formation is not considered while the bottom panels show the same in the presence of
rope hadronization.}
\label{fig7TeV2d}
\end{center}
\end{figure*}

\begin{figure*}
\begin{center}
\includegraphics[scale=0.78]{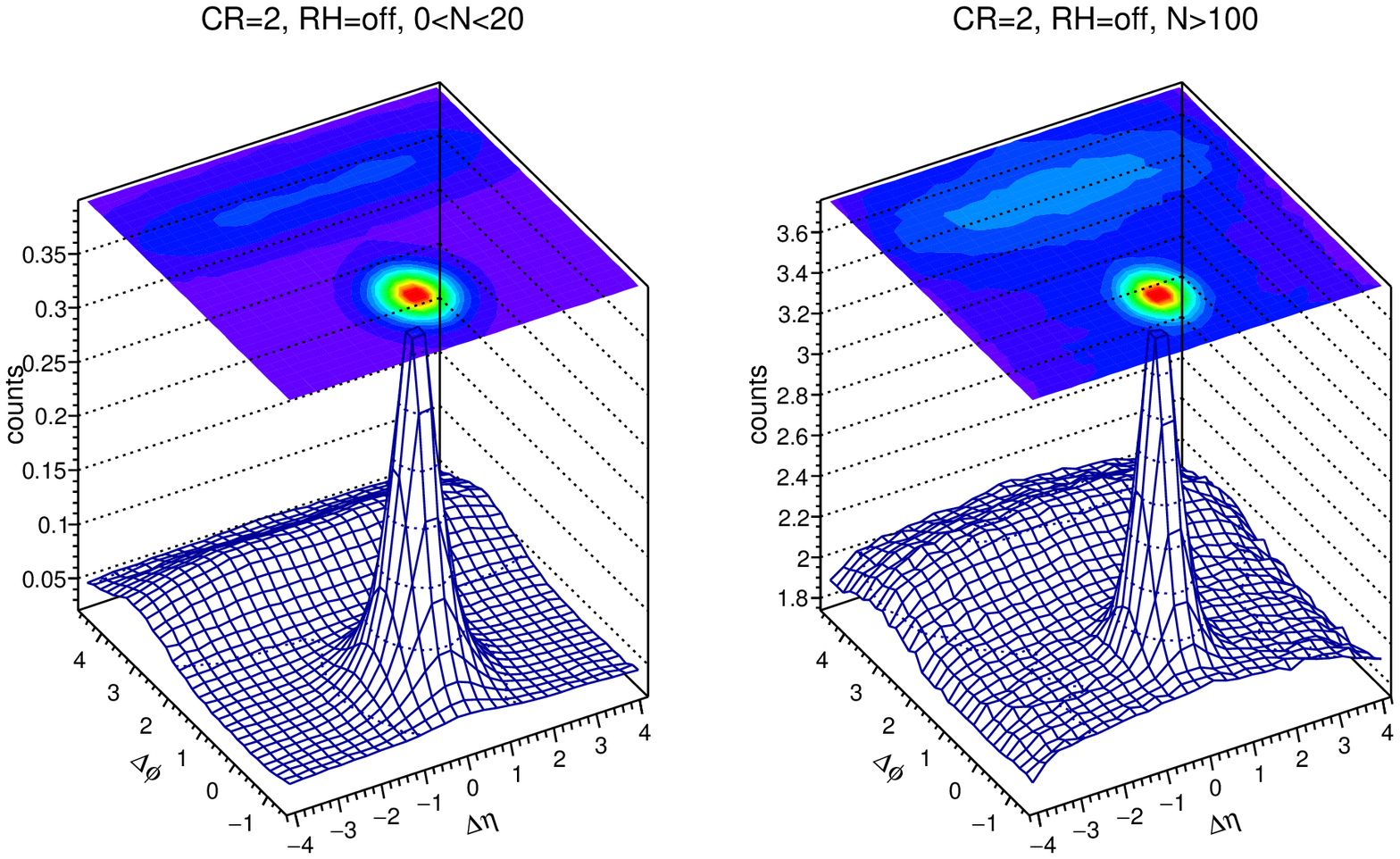}
\includegraphics[scale=0.78]{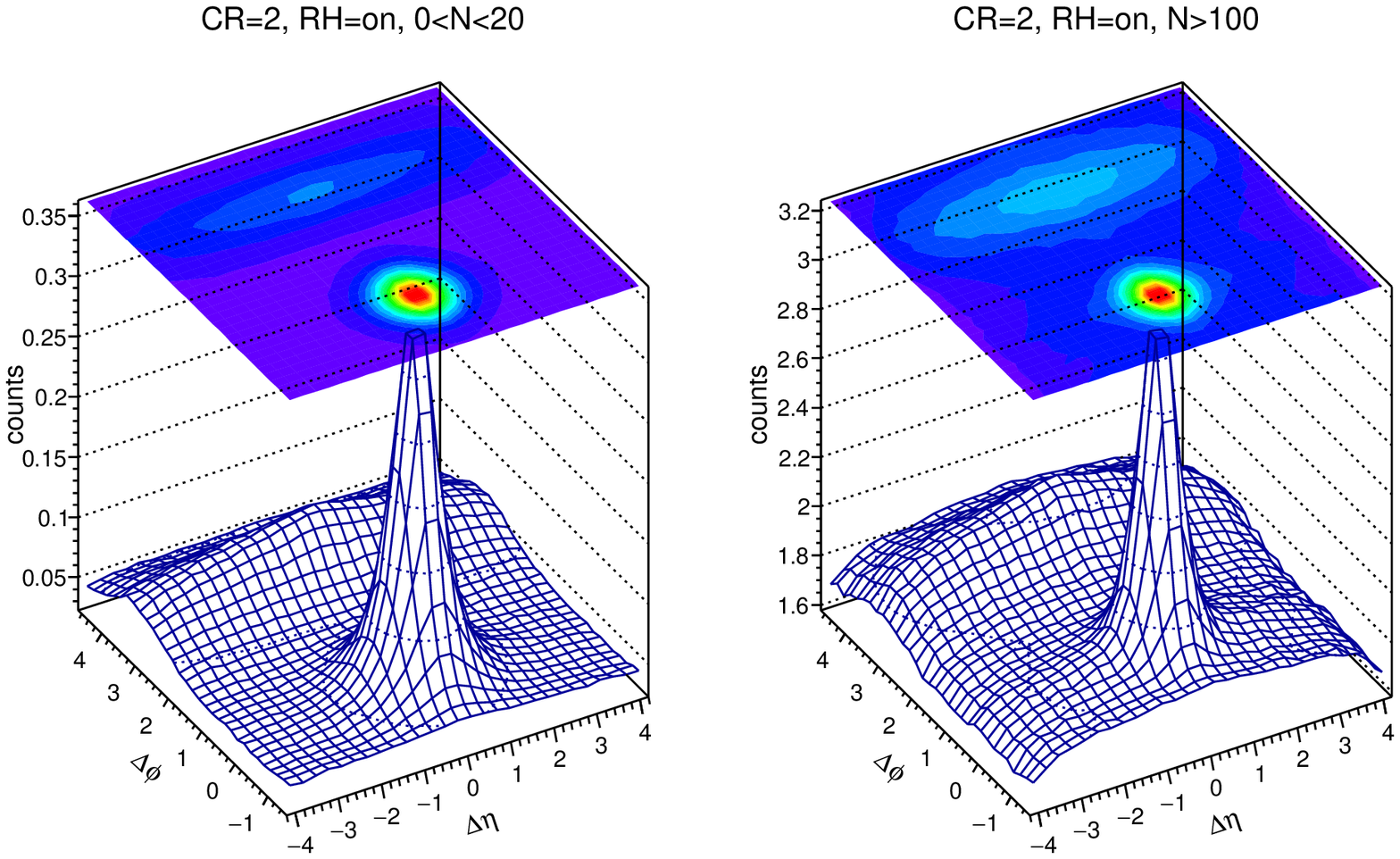}
\caption{(Color online) The two-dimensional $\Delta\eta-\Delta\phi$ correlation function of charged particle pairs in p$-$p collisions at  $\sqrt{s}$ = 13 TeV for two different multiplicity classes with $1.0 < p_{T}^{Trigg}<
3.0$  GeV/c and  $1<p_{T}(Asso)< 3.0$. The top panels refer to the case when rope formation is not considered while the bottom panels show the same in the presence of
rope hadronization.}
\label{fig13TeV2d}
\end{center}
\end{figure*}

\begin{figure*}
\begin{center}
\includegraphics[scale=0.8]{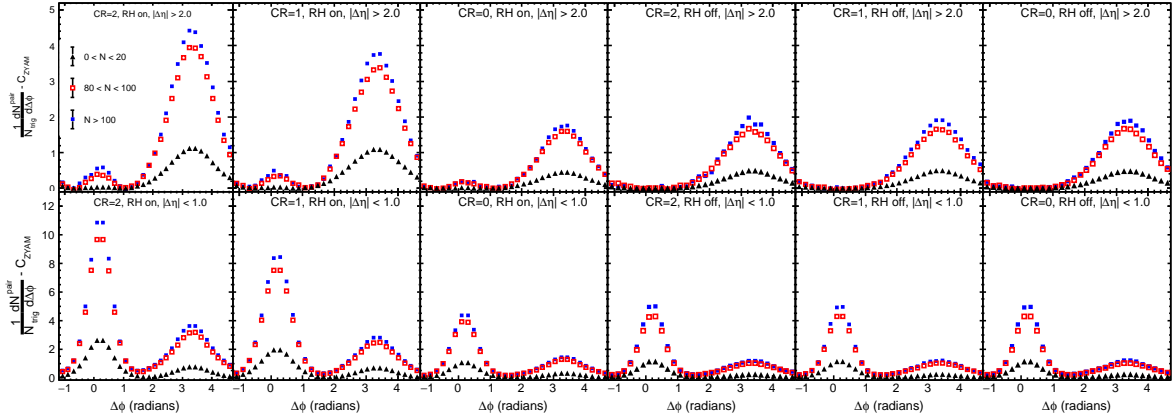}
\caption{(Color online) The one-dimensional  $\Delta\phi$ projection of two particle correlation function of charged particles 
in  p$-$p collisions at $\sqrt{s}$ = 13 TeV for three different multiplicity  classes. The top panels refer to the long range region ($\Delta\eta  > 2.0$) while the lower ones refer to the short range region ($\Delta\eta  < 1.0$).}
\label{fig13TeV}
\end{center}
\end{figure*}

The two-dimensional  $\Delta\eta-\Delta\phi$ correlation functions for charged particle pairs (charged trigger particle and charged associated particle) are  
shown for two different multiplicity classes in p$-$p collisions at  $\sqrt{s}$ = 7 TeV and 13 TeV in Figure \ref{fig7TeV2d} and  Figure \ref{fig13TeV2d}, respectively.
The figures also compare the correlation function of the charged particles for the effect of rope hadronization in presence of color reconnections for the considered multiplicity 
classes. An expected correlation peak near $\Delta\eta-\Delta\phi$ = (0,0) is observed for both low and high multiplicity class for both the
energies. The peak primarily originates due to jet fragmentation as mentioned before. An away side ($\Delta\phi \sim \pi$) ridge-like structure extended up to higher values of 
$|\Delta\eta|$ (containing the contributions from back-to-back jets) is also visible in all the figures. 
However, a long-range ridge-like structure extending almost  three units in $\Delta\eta$  can be seen in the near side ($\Delta\phi \sim 0$) region of the correlation function for high multiplicity class where the formation of ropes is more pronounced. This structure is not observed in low multiplicity class where the formation of ropes is less probable due to lack of overlapping strings.  Moreover, the structure is not present in higher multiplicity class when the formation of ropes in not considered. This ridge-like structure is quite similar to the observed ridge in high multiplicity p$-$p collisions. The details of the observed two-dimensional correlation functions were further investigated by projecting the event-averaged distributions into one-dimensional distribution in $\Delta\phi$ for different $\Delta\eta$ ranges. The projections over different $\Delta\eta$ range were divided into two groups, namely the short range (when projected over $|\Delta\eta| < 1.0$, the jet region) and the long range ($|\Delta\eta| > 2.0 $, the ridge region). The standard zero-yield-at-minimum (ZYAM) method is employed to estimate the associated yield in the correlated region \cite{zyam}. In this method, the one-dimensional  $\Delta\phi$ correlation distribution is fitted with a second order polynomial function in the range of $0.1 < |\Delta\phi| < 2 $.  The lowest value of the fitted polynomial function, $C_{ZYAM}$  (which accounts for the constant pedestal) is subtracted from the distribution.  As a result, the minimum of the correlation function has zero associated yield. This procedure is implemented to correct the one-dimensional $\Delta\phi$ correlation function in  p$-$p collisions at both the energies for both long and short-range region. This study was performed for three different multiplicity classes with three different modes of color reconnections within (and without) the framework of rope hadronisation. 
Figure  \ref{fig13TeV} compares the effect of rope hadronisation in one-dimensional $\Delta\phi$ correlations for three different modes of color reconnections in p$-$p collisions at 13 TeV.  
The top panels which focus on long range region show an expected 
away-side peak emanating from back-to-back jets for all the three multiplicity classes. However, a non-zero associated yield peak is also 
observed in the near-side for long-range region for high multiplicity events when the rope hadronization is enabled. This peak is absent for 
low multiplicity events as well as for events without rope hadronization for both the energies. This observation indicates that the microscopic phenomenon of partonic color reconnections together with the formation of ropes due to overlapping strings followed by shoving and subsequent hadronisation in high multiplicity events can also generate ridge-like structure in the near side region. The bottom panels of the figure show the short range region and one can observe the expected near side peak originating due to jets and an away-side ridge-like structure (due to back-to back jets and momentum conservation). The strength of the correlation function is higher for CR-1 and CR-2 mode compared to CR-0 mode. This may be attributed to additional reconnections in the former modes due to junction formation and shifting the gluon position in order to minimise the string length. A detailed study with identified particles is required to gain more understanding. Nevertheless, the observed ridge-like structure in the near-side region is  qualitatively similar to the ones observed in data  which further supports the idea that the microscopic processes at partonic level can mimic collectivity like features without assuming the formation of a thermalised medium.  

\section{Summary}
The observation of long range $\Delta\eta-\Delta\phi$ correlations of charged particles at near-side ($\Delta\phi \sim 0$) measured by LHC experiments 
in high multiplicity p$-$p collisions indicated towards the presence of small scale collective effects which are similar 
to that observed in p$-$A (nucleon-nucleus) and A$-$A (nucleus-nucleus) collisions. This novel
observation was investigated by studying the two particle correlation between charged particles in p$-$p collisions
at $\sqrt{s}$ = 7 TeV  and 13 TeV  using Pythia 8 event generator within the framework of final-state partonic 
color reconnection mechanism as well as the microscopic rope hadronization model.  The formation of ropes due to overlapping of
strings followed by string shoving due to the creation of pressure gradient in the overlap region could generate the long range correlations in the near-side 
region of $\Delta\eta-\Delta\phi$ distribution for high multiplicity events in p$-$p collisions.  This near-side 
ridge-like structure observed in the long range region is qualitatively similar to the ones measured in high multiplicity p$-$p collisions at LHC and thus can 
provide an alternate view of ridge formation when the formation of a thermalised medium is not imperative.

\section{Acknowledgements}
The authors would like to thank the Department of Science and Technology (DST), India for supporting the present work.

\end{document}